\documentclass[twocolumn,showpacs,preprintnumbers,amsmath,amssymb]{revtex4}
\usepackage{graphicx}

\begin{document}
 %\preprint{Draft}
   \title{Comments on ``Modeling Galaxy Halos Using Dark Matter with Pressure''}
   \author{Kung-Yi Su$^{1,3}$}
     \email{b95202049@ntu.edu.tw}
   \author{Pisin Chen$^{1,2,3,4}$}
     \email{pisinchen@phys.ntu.edu.tw}
     \affiliation{%
1. Department of Physics, National Taiwan University, Taipei, Taiwan 10617\\
2. Graduate Institute of Astrophysics, National Taiwan University, Taipei, Taiwan 10617\\
3. Leung Center for Cosmology and Particle Astrophysics, National Taiwan University, Taipei, Taiwan 10617\\
4. Kavli Institute for Particle Astrophysics and Cosmology, SLAC National Accelerator Laboratory, Stanford University, Stanford, CA 94305, U.S.A.
}%

   \date{\today}

\begin{abstract}
We comment on the calculational mistake in the paper ``Modeling galaxy halos using dark matter with pressure'' by Somnath Bharadwaj and Sayan Kar. The authors made a mistake while calculating the metric, which led to an overestimate of the deflection angle of light passing through the halos for $-1<w _r<-0.5$ and an underestimate of the deflection angle for $-0.5<w _r<0$. In addition, solution for $w _r>0$ should not exist. Although the Bharadwaj-Kar solution should be corrected, it appears that the characteristics of the deflection angle under the supposed non-conventional non-ideal fluid equation of state for the dark matter halo remain sensitive to the impact parameter and may be verifiable through observations.
\end{abstract}

\maketitle

In the paper ``Modeling galaxy halos using dark matter with pressure''\cite{Bharadwaj}, Somnath Bharadwaj and Sayan Kar inspected the possibility that the dark matter could be a non-ideal fluid with a significant pressure. They showed that such a non-conventional notion of dark matter could be verified through the observed galactic rotation curves and the modification on the gravitational lensing deflection angle. This is interesting and different from the conventional approach to dark matter. Such a dark matter cannot be accommodated by the Newtonian gravity and it has to resort to general relativity. They model the galactic dark matter halo based on the following assumptions:
\begin{enumerate}
\item
The metric inside the halo is spherically symmetric and the proper time interval $d\tau$ is
  \begin{eqnarray}
c^2d\tau^2=e^{2\Phi(r)}c^2dt^2-e^{2\lambda(r)}c^2dr^2-\-r^2d\Omega^2
\;,
   \label{eq:one}
  \end{eqnarray}
 and it reduces to Schwarzschild metric outside the boundary, $r=R$, of the halo.
 \item
 The gravitation is weak.
 \item
 The pressure inside the halo is anisotropic, which means that the radial pressure $P_r$ is different from the tangential pressure $P_t$.
 \item
 The equation of state is either $P_r=w _r\rho c^2$ or $P_t=w _t\rho c^2$, where $w_r$ and $w_t$ are constants and $\rho$ is the energy density.
\end{enumerate}

\subsection{Metric}

\begin{table*}
%\caption{\label{tab:table1}Bharadwaj and Kar's solution}
   \begin{center}
TABLE I. Bharadwaj-Kar solution   ($r<R$)
   \end{center}
  \begin{ruledtabular}
  \begin{tabular}{cccc}
$w_r$ &$\lambda(r)=\lambda_N\times$ & $\rho(r)=\rho_N\times$  & $P_t(r)=(c^2/2)\rho_N(r)\times$\\
\hline
$-1<w_r<\infty$ & $\frac{1}{1+w_r}[1+w_r(\frac{r}{R})^{-\frac{w_r}{1+w_r}}]$ & $\frac{1}{1+w_r}[1+\frac{w_r}{1+w_r}(\frac{r}{R})^{-\frac{w_r}{1+w_r}}]$ &$(\frac{w_r}{1+w_r})^2(\frac{r}{R})^{-\frac{w_r}{1+w_r}}$ \\
$-1$ &$1-\ln(r/R)$ & $\ln(r/R)$&1\\
  \end{tabular}
  \end{ruledtabular}
%\caption{\label{tab:table1}The correct solution}
 \begin{center}
TABLE II. The corrected solution ($r<R$)
 \end{center}
  \begin{ruledtabular}
  \begin{tabular}{cccc}
$w_r$ &$\lambda(r)=\lambda_N\times$ & $\rho(r)=\rho_N\times$  & $P_t(r)=(c^2/2)\rho_N(r)\times$\\
\hline
$-1<w_r<0$ & $\frac{1}{1+w_r}[1+w_r(\frac{r}{R})^{-\frac{1+w_r}{w_r}}]$ & $\frac{1}{1+w_r}[1-(\frac{r}{R})^{-\frac{1+w_r}{w_r}}]$ &$(\frac{r}{R})^{-\frac{1+w_r}{w_r}}$ \\
$-1$ &$1-\ln(r/R)$ & $\ln(r/R)$&1\\
  \end{tabular}
  \end{ruledtabular}
\end{table*}

  The task is to determine $g_{00}$ and $g_{rr}$, or $\Phi(r)$ and $\lambda(r)$, in the metric.  $\Phi(r)$, and therefore $g_{00}$, can be fully determined by the flat rotation curve. $\lambda(r)$, on the other hand, can be determined by the equation of state with the aid of the Einstein equations. After substituting the $\Phi(r)$, the Einstein equations lead to three equations for $\lambda(r)$:
  \begin{eqnarray}
\frac{(r\lambda)'}{r^2}&=&\frac{4\pi G}{c^2}\rho\, \cr
\frac{(v_c/c)^2-\lambda}{r^2}&=&\frac{4\pi G}{c^4}P_r\, \cr
-\frac{\lambda'}{r}&=&\frac{8\pi G}{c^4}P_t\; ,
   \label{eq:one}
  \end{eqnarray}
where $v_c\sim$ 200 km/s is the constant rotation velocity of the halo associated with the flat rotation curve of our galaxy. Using the above metric, one finds that the deflection angle depends strongly on the equation of state. Accordingly with the aid of the observed deflection angle, the equation of state can be determined. Conversely, if the equation of state of the dark matter halo is known, one can in principle test the general relativity at galactic scale.

While we agree with the framework of their approach, we find some mistakes in their calculation of the metric under the equation of state $P_r=w _r\rho c^2$, which result in an overestimate of the deflection angle of light passing through the halo for $-1<w _r<-0.5$ and an underestimate of it for $-0.5<w _r<0$.  In addition, solution for $w _r>0$ should not exist. On the other hand, we found no mistake in their calculation under the condition $P_t=w _t\rho c^2$, and the corresponding conclusions for that case remain unchanged.

Their original solutions and our corrected expressions of the metric, the density and the pressure are displayed in Table 1 and Table 2, respectively, for comparison. Note that $\lambda_N=(v_c/c)^2$ and $\rho_N=v_c^2/4\pi Gr^2$.

It can be seen that the main difference between the two results lies in the second term of $\lambda$, where the exponent should be $-(1+w_r)/w_r$ instead of $-w_r/(1+w_r)$. As a consequence, the acceptable range of $w_r$ should be $[-1,0]$ rather than $[-1,\infty)$, so as to ensure the positive definiteness of the energy density. It should be evident that the two solutions are identical when $w_r=-0.5$. Even so, several qualitative features of the Bharadwaj-Kar solution still hold in our results. For example, for $P_r=-\rho c^2$ the metric elements satisfy the relation $g_{11}=-g_{00}^{-1}$ inside and outside the halo in both calculations.

Bharadwaj and Kar commented in their paper that $P_r=-w_r\rho  c^2$ is the only condition for $\rho=0$ at $r=R$. They also mentioned that in general the pressure and the energy cannot be matched at the halo boundary and would therefore be discontinuous. However we see from the corrected solution that the situation is not so bleak. For the equation of state $P_r=w _r\rho c^2$, we find that both the radial pressure and the energy density are 0 and continuous at the boundary, although the tangential pressure remains discontinuous. As for the equation of state $P_t=w _t\rho c^2$, the radial pressure is 0 and continuous at the boundary but the energy density and tangential pressure are not.

\subsection{Deflection Angle}

The general formula for the deflection angle in the weak field limit, i.e., Eq(15) in the original paper, is
 \begin{eqnarray}
\delta=-2\Big(\frac{c}{v_c}\Big)^2\int_{0}^{1/b} \frac{d\alpha}{dy}\Big(\frac{1}{b^2}-y^2\Big)^{-1/2} \mathrm{d}y\, ,
\;
   \label{eq:one}
 \end{eqnarray}
where $\delta$  is the deflection angle evaluated in units of $(v_c/c)^2$, $y=u+\alpha(y)$, $\alpha(y)=[(\Phi+\lambda)/b^2-\lambda y^2]/y$ and $u=1/r$. $b$ is the impact parameter that equals to $L/E$, where $L=r^2(d\Phi/d\tau)$ and $E=e^{2\Phi}c(dt/d\tau)$ are the constants of motion of the system. For a given equation of state, the deflection angle can be obtained based on the above formula. Fig.1 and Fig.2 display the author's original and our corrected results, respectively, of the deflection angle as a function of $b/R$ under the equation of state  $P_r=w _r\rho c^2$.

\begin{figure}
\centering
\includegraphics[width=8cm]{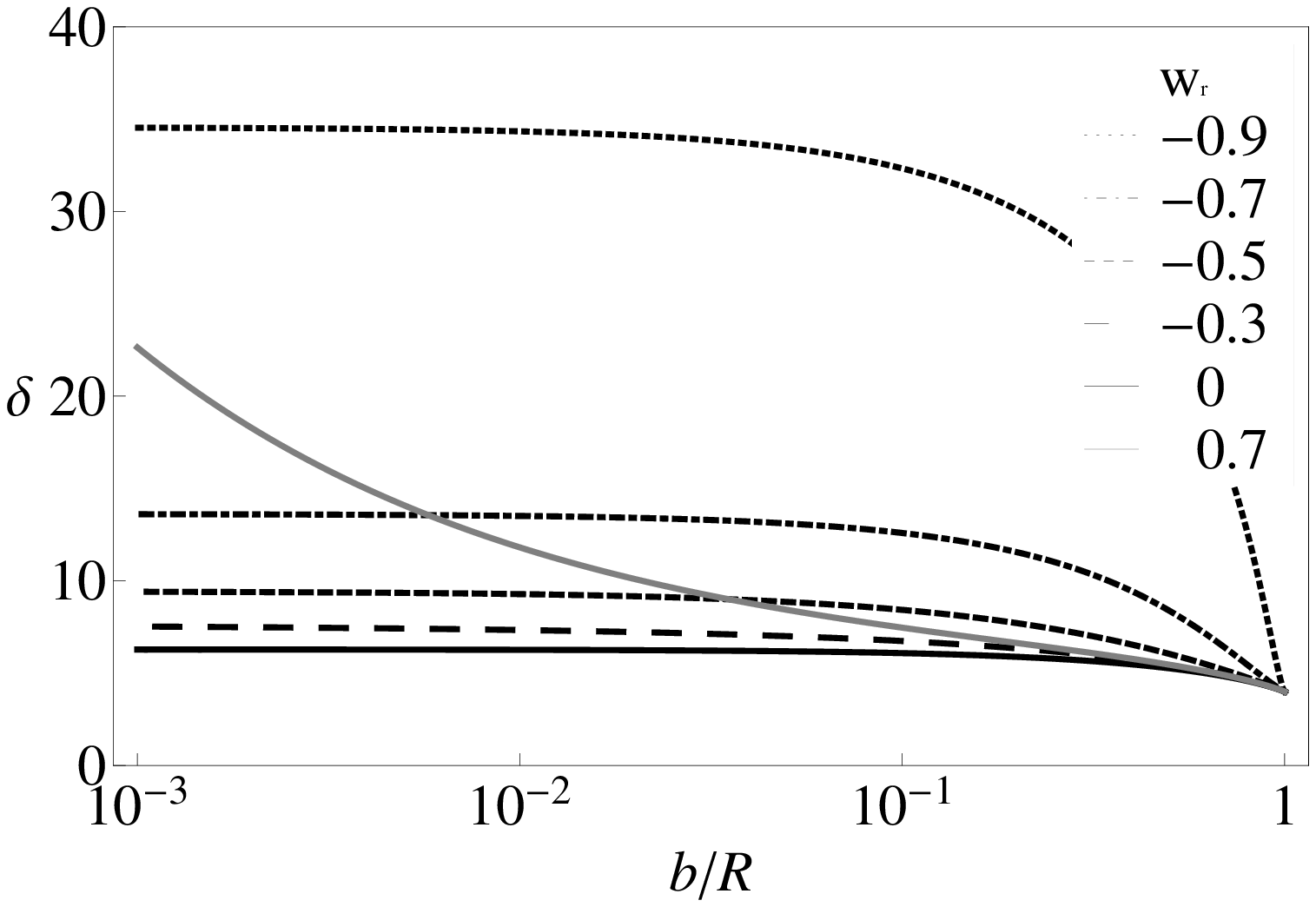}
  \caption{\label{fig:epsart} The deflection angle under $P_r=w _r\rho c^2$ based on the Bharadwaj-Kar solution.}
\includegraphics[width=8cm]{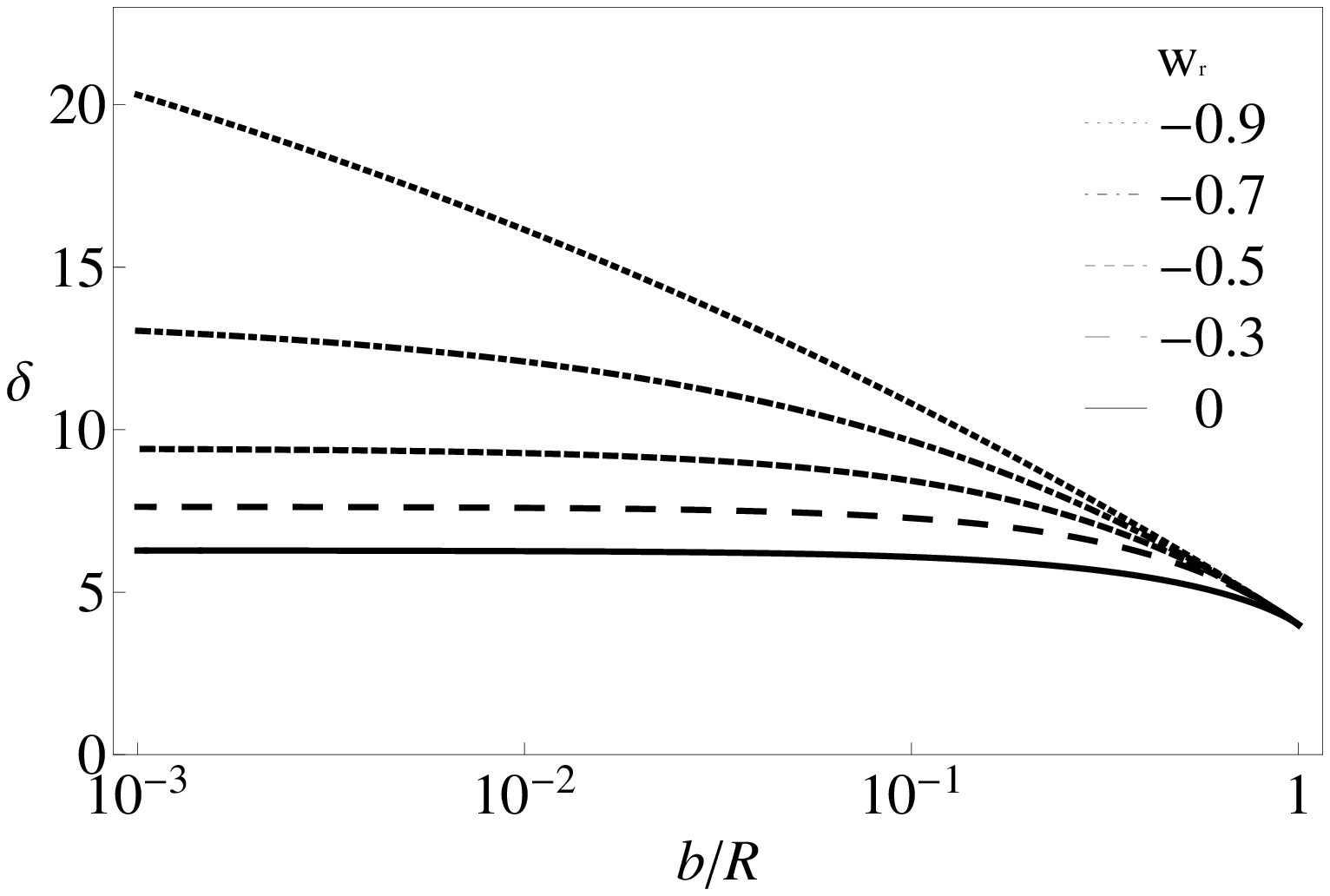}
  \caption{\label{fig:epsart} The corrected deflection angle under $P_r=w _r\rho c^2$.}
\end{figure}

For the case $P_r=w _r\rho c^2$, several features of the deflection angle remain unchanged in both solutions. When $w_r<0$, the deflection angle is always larger than that in the case where there is no pressure. Furthermore, the deflection angle increases as $b/R$ decreases, and it approaches a constant that is larger than $\delta_N=2\pi$, which is the deflection angle for an infinite halo with no pressure. The main difference, however, is that the corrected deflection angle converges more slowly to a smaller value for $w_r<-0.5$, while faster to a larger value for $w_r>-0.5$. We note that in Fig.2 the deflection angle for $w_r=-0.9$ in our calculation does converge, though very slowly. In addition, we see from the figures that there exists a ``turning point '' for some of the curves. In the Bharadwaj-Kar solution, the turning point is more visible for larger values of $w_r$, while it is more visible for smaller values in our calculation.

In conclusion, our corrected solution for the case $P_r=w _r\rho c^2$ leads to smoother energy and pressure profiles at the halo boundary. In addition, the corresponding deflection angle converges more slowly to a smaller value for $-1<w _r<-0.5$, while it converges more quickly  to a larger value for $-0.5<w_r<0$. The good news is, under the new solution the deflection angle remains sensitive to the equation of state. So it may still be possible to verify the proposed non-conventional, non-ideal fluid equation of state for the dark matter halo through observations\cite{Faber}.


\begin{thebibliography}{99}

\bibitem{Bharadwaj}
S. Bharadwaj and S. Kar, {\sl Phys. Rev. D} {\bf 68}, 023516 (2003).

\bibitem{Faber}
T. Faber and M. Visser, {\sl Mon. Not. R. Astron. Soc.} {\bf 372}, 136 (2006).

\end{thebibliography}
\end{document}